\documentclass{JINST}
\usepackage{graphicx}
\usepackage{fancyvrb}
\usepackage{url}
\usepackage{lineno}
\DefineVerbatimEnvironment{code}{Verbatim}{fontsize=\small}

\title{Performance of a cryogenic system prototype for the XENON1T Detector}

\author{E.~Aprile$^a$, R.~Budnik$^a$\thanks{Corresponding author. E-mail: \texttt{ranny@astro.columbia.edu}}, B.~Choi$^a$, H.~A.~Contreras$^a$, K.~L.~Giboni$^{a}$, L.~W.~Goetzke$^a$, R.~F.~Lang$^{ab}$, K.~E.~Lim$^a$, A.~J.~Melgarejo$^a$, G.~Plante$^a$, A.~Rizzo$^a$, P.~Shagin$^c$ \\
\llap{$^a$}Physics Department, Columbia University, New York, NY 10027, USA \\
\llap{$^b$}Current position: Department of Physics, Purdue University, West Lafayette, IN 47907, USA\\
\llap{$^c$}Department of Physics and Astronomy, Rice University, Houston, TX 77005 - 1892, USA}

\abstract{We have developed an efficient cryogenic system with heat exchange  and associated gas purification system as a prototype for the XENON1T experiment. The XENON1T detector will use about 3 tons of liquid xenon (LXe) at a temperature of 175K as target and detection medium for a dark matter search.  In this paper we report results on the cryogenic system performance  focusing on the dynamics of the gas circulation-purification through a heated getter, at flow rates above 50 Standard Liter per Minute (SLPM). A maximum flow of 114 SLPM has been achieved, and using two heat exchangers in series, a heat exchange efficiency better than 96\% has been measured.}

\keywords{Liquid xenon; Cryogenics; Cooling power; Gas purification; Heat Exchange}

\begin{document}

\section{Introduction}

The worldwide race towards direct dark matter detection in the form of
Weakly Interacting Massive Particles (WIMPs) has been dramatically accelerated by the remarkable progress and evolution of liquid xenon time projection chambers
(LXeTPCs). The XENON100 experiment has already placed the most stringent limits on the WIMP-nucleon spin-independent cross section \cite{Aprile:2011xe} and new data have been accrued at the Laboratori Nazionali del Gran Sasso (LNGS) in Italy towards its ultimate sensitivity goal of $2\times 10^{-45}$~cm$^2$. The next phase of the XENON program will use a LXeTPC with about 3  tons of LXe \cite{Xe1t}.

To demonstrate some of the technologies relevant to the realization of such a massive LXe detector, a system was developed and tested. This system, the XENON1T Demonstrator, consists of a LXeTPC, a cryocooler, a gas system with recirculation pump and hot getter, and a heat exchanger (HE) module. The R$\&$D results presented here are especially relevant for addressing the key requirement of ultra-high purity LXe, enabling free electrons to drift over distances larger than 1~m and long scintillation photon absorption lengths.  The $\sim 3$ tons of Xe filling XENON1T must contain less than a ppb (part per billion) level of $\mathrm{O_2}$-equivalent electronegative impurities  ~\cite{Schmidt:2001},  
and  $\mathrm{H_2O}$  at a similar low level ~\cite{Baldini2005}. 

Materials outgassing  is a constant source of electronegative impurities. Although the detector vessel and TPC components can be baked-out in order to accelerate the outgassing, residual outgassing remains  and necessitates continuous purification throughout the operation of the experiment. The gas recirculation rate in the XENON100 experiment, filled with 161 kg of LXe, was limited to 5~SLPM by the available cooling power~\cite{Aprile:2011dd}. Nevertheless, this resulted in a significant decrease in electron attenuation over months of operation time, reaching attenuation lengths $>$~1~m. For XENON1T, we plan to increase the recirculation speed to around 100~SLPM, which corresponds to about 800~kg/day. This is required to enable a reasonable commissioning period of the detector with good performance in terms of charge and light yields. The dynamics of gaseous xenon at this speed causes large pressure gradients and requires components that can handle that flow, including tubing, recirculation pump, getter and flow controller.

Xenon gas flow rates  in excess of 40~SLPM through a purification system based on a hot getter have not been reported in the literature to-date. While most components used in the cryogenic and gas systems developed for the XENON1T R\&D are commercially available, their suitability for Xe gas must be proven. As an example, the performance of commercially available hot metal getters, such as SAES Monotorr \cite{SAES}, used to remove  electronegative impurities from noble gases, is typically tested with argon gas.  The impact of the larger xenon density and heat capacity  on the purification efficiency and flow characteristics of the getter must be studied.

Membrane-based gas circulation pumps are commercially available as well, but the high density of xenon gas limits the ability of these pumps to work at high flow rates over long time periods, mostly due to the wear of the diaphragms from high pressures  and induced heating. In addition, their leak tightness and durability under these conditions must be investigated. 

Finally, since the recirculation and purification is done in the gas phase, the Xe gas must be continually re-liquefied, requiring large amounts of available cooling power. 
To cool down Xe gas at  a rate of 1~SLPM from room temperature to 175~K, 
less than 2~W are used, out of a total of about 10.6~W that are needed to liquefy and cool at the same rate. 

At 100~SLPM, this translates to more than 1~kW of cooling power, which is not practical as the overall efficiency of the cryocoolers successfully tested within the XENON program is limited due to the high power consumption of the cooling system.  An efficient heat exchange to reduce the cooling power cost of compensating the evaporation and heating of Xe gas is essential. The efficiency of a commercial parallel-plate HE, to transfer the heat within the system and use it to cool purified Xe gas before injecting it back into a LXe detector, has already been studied as reported in~\cite{Giboni11}. Here we extend these studies to the high flow rate regime.

\section{Experimental Setup}
The XENON1T Demonstrator apparatus, constructed and tested at Columbia University, consists of a liquid xenon detector, a cooling tower with a cryocooler, a gas recirculation system with a pump and a  getter, and a HE module. The detector, cooling tower and  HE are mounted in three separate vacuum-insulated vessels to reduce heat losses to the ambient air and are super-insulated with 12 layers of aluminized mylar. The cryocooler, an Iwatani PC-150 Pulse Tube Refrigerator (PTR) with a 6.5 ~KW water-cooled He compressor, delivers  200~W of cooling power at 165K. This PTR is the same as used on XENON100. A HE is used to cool and liquefy xenon gas returning from the hot getter. The liquid is taken from the detector through the HE, where the latent heat is transferred to the returning xenon gas stream with an  efficiency greater than 96\% (e.g. \cite{Giboni11}).

For the measurements reported here, the detector was not implemented as a TPC, but was merely used as a double walled vacuum insulated vessel. A maximum of $\sim$7~liters (about 20~kg) of LXe when filled for these tests.

\subsection{Gas purification system}
\begin{figure}[htb]
\begin{center}
\includegraphics[width=10cm]{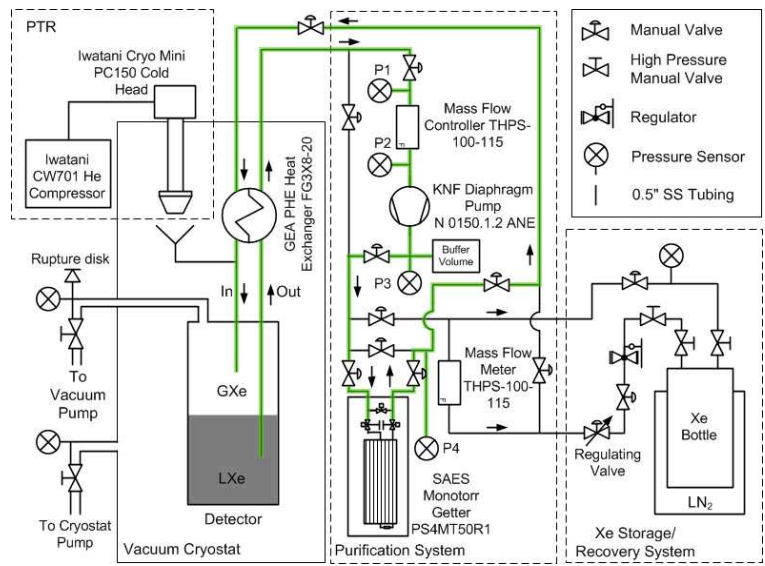}
\caption{Schematic diagram of the gas system  for the XENON1T demonstrator. The green line follows the recirculation path through the gas system.}
\label{fig:demo_gas_sys}
\end{center}
\end{figure}

The gas system was specifically designed for high speed closed-loop circulation of the Xe gas. The lines are made of 1/2'' stainless steel pipes with VCR fittings, to allow the rapid flow through the circulation loop.  Figure \ref{fig:demo_gas_sys} shows a schematic of the gas system, including all the equipment used in the setup. The gas purifier is a SAES Monotorr getter, model PS4MT50R1, rated for purification of rare gases at flows up to 75 SLPM~\cite{SAES}. 

A large capacity double-headed diaphragm pump (KNF 1400 series) was selected, nominally capable of flowing $ \sim 200 $ SLPM of air at atmospheric pressure on the input. The pump was modified from double diaphragm to single diaphragm (for each of the two heads) to accommodate our requirement to withstand an output pressure greater than $6$ bar. Water cooling was also added to mitigate heating of the pump heads. The flow is controlled with a Teledyne Hastings Mass Flow Controller (MFC) Model HFC-303 \cite{MFC_man}, calibrated up to 250 ~SLPM of Xe gas. The control valve is situated at the inlet of the pump, since limiting the flow  on the outlet may result in too high pressure (10 bars and above) which may damage the pump diaphragms. A buffer volume of about one liter was installed at the outlet of the pump, in order to damp pressure fluctuations.

Four pressure gauges were mounted at different positions along the recirculation loop. $ P_1 $ measures the pressure at the input of the MFC and the output of the HE, $ P_2 $ measures the pressure at the input of the pump and the output of the MFC, $ P_3 $ measures the pressure at the output of the pump and input of the getter and $ P_4 $ measures the pressure at the output of the getter, which is also the pressure going back into the HE. The pressure gradients give an indication of the resistance to the flow, allowing the system design to be optimized for high flow rates.

\subsection{Cryogenics and Heat Exchanger}
\begin{figure}[htb]
\begin{center}
\includegraphics[width=6cm]{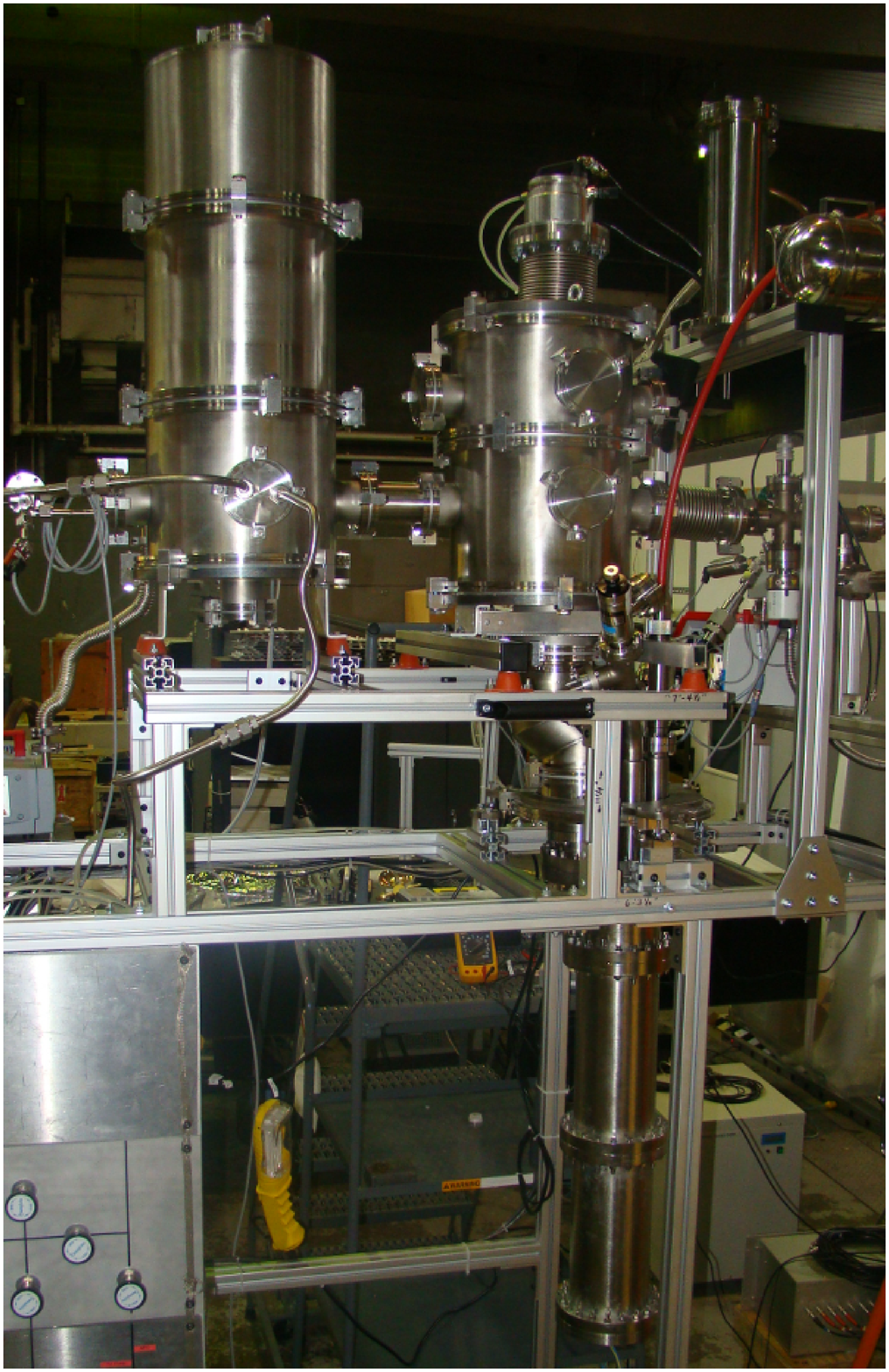}
\includegraphics[width=6cm]{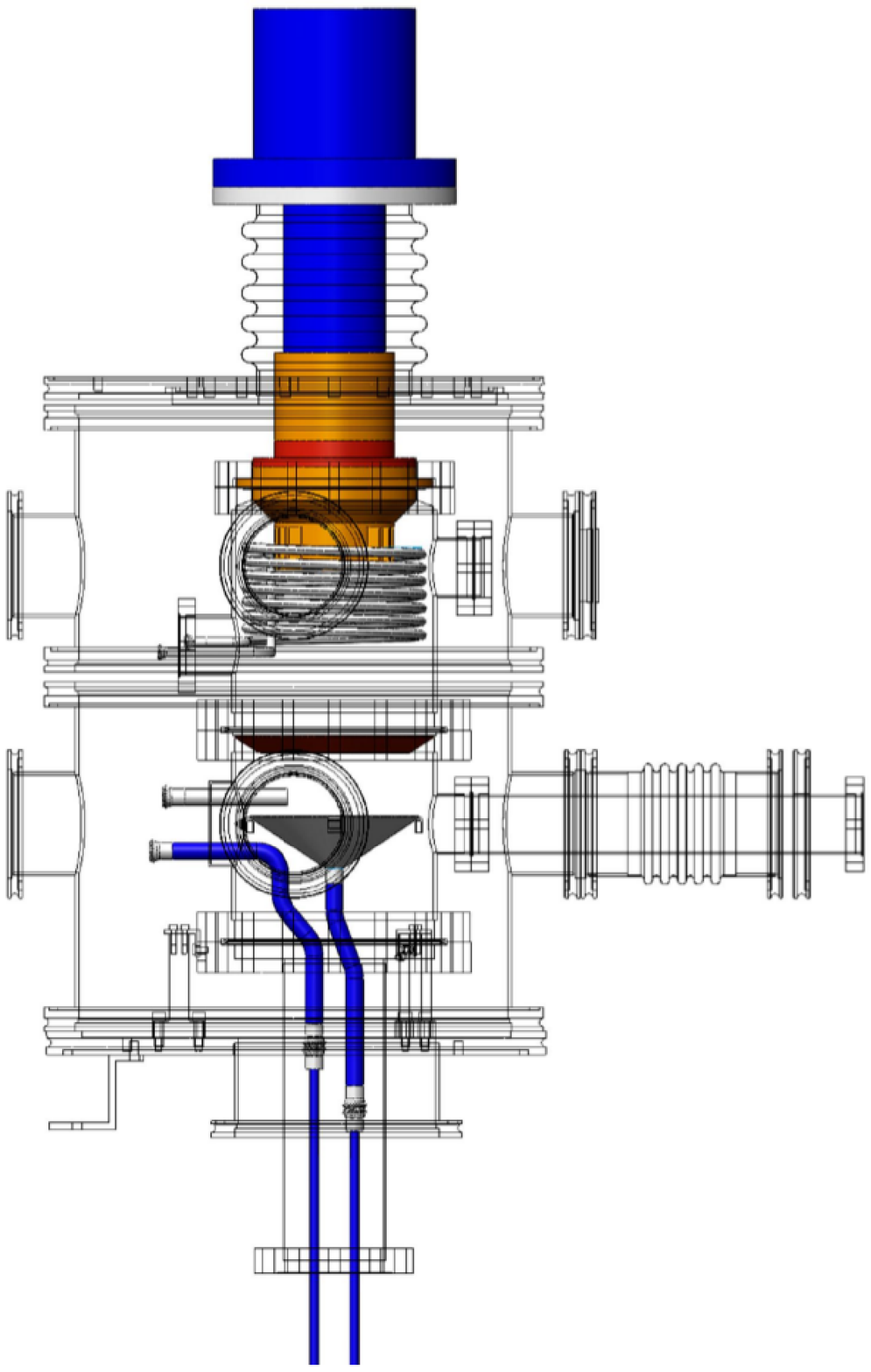}
\caption{(Left): The XENON1T Demonstrator setup at Columbia Nevis Laboratory. \newline  (Right): A technical drawing of XENON1T Demonstrator cooling tower with insulation jacket and the PTR.}
\label{fig:demo}
\end{center}
\end{figure}

Figure \ref{fig:demo} shows a photograph and a CAD drawing of the experimental setup. 
The cooling power of the PTR is delivered to the Xe system through a copper cold finger at the top of the cooling tower. There, Xe liquefies on the fins of the cold finger, drips down and is collected by a funnel before flowing into the LXe chamber, located below the cooling tower. The entire system is surrounded by an insulation vacuum to minimize heat leaks into the system, as well as layered aluminized mylar for super-insulation.  Due to the narrow temperature margin of only 3.4~K between the liquid and solid phase of Xe (under atmospheric pressure), temperature control is especially important.  For temperature control, a copper cup with electrical heaters is inserted between the cold head of the PTR and the cold finger that reaches into the detector volume. The maximum power of these heaters is limited to the cooling power of the PTR as a fail-safe solution to prevent overheating. The temperature of the assembly above and below these heaters is measured with LakeShore PT111 resistors and monitored continuously. A Lakeshore 340 Proportional, Integral and Differential (PID) controller reads the temperature at the cold finger and controls the power to the heaters. The heaters keep the temperature of the cold finger at the set value and thus provide a constant cooling power. The temperature of the liquid is stable to better than 0.04~K over extended periods of continuous operation of the system.

Xe purification is done by continuous gas circulation through the  hot getter, taking LXe from the detector
into the HE and letting the returning Xe gas cool and liquefy inside the HE on its way back into the detector.
The HE module is located in a separate vacuum insulated vessel.

\begin{figure}[htb]
\begin{center}
\includegraphics[width=6cm]{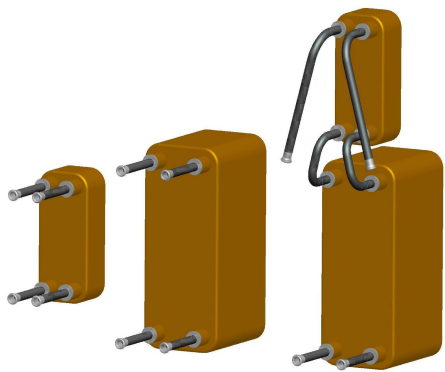}
\caption{The three configurations of HEs tested in this work. The leftmost is configuration a, the middle is configuration b and the rightmost is configuration c (see text).}
\label{fig:HE_setups}
\end{center}
\end{figure}

The heat exchange was achieved with parallel plate HEs available for commercial use, and already tested in \cite{Giboni11} . Two HEs of different size, both made by GEA  \cite{GEA} were used. The smaller unit consists of 20 plates (model FG3X8-20, measuring $3.3 \times 7.8 \times 2.1$ inches, with a volume of about 0.5 l) and the larger of 60 plates (model FG5X12-60, measuring $4.9 \times 12.2 \times 6$ inches, with a volume of about 3.8~l). Three configurations of HEs were tested: a) the small HE only, b) the large HE only, c) a combination in which the two HEs are connected in series, with the larger one at the bottom  connected to the LXe side (the detector) and the smaller one, on top,  connected to the gas system. A schematic diagram showing the three configurations is shown in Figure \ref{fig:HE_setups}.

\section{Results and Discussion}

\subsection{Fast gas circulation}\label{subsec:gas_circ}
In this section we describe the results of the dynamics of fast recirculation flow of Xe gas, regardless of cryogenics. The flow rate is constant through the circulation path, and its self consistency  is maintained by pressure differences between different points along the path. The pressure differences drive the flow against the restrictions (resistances) of the components - pipes, valves, getter etc.

\begin{figure}[htb]
\begin{center}
\includegraphics[width=8cm]{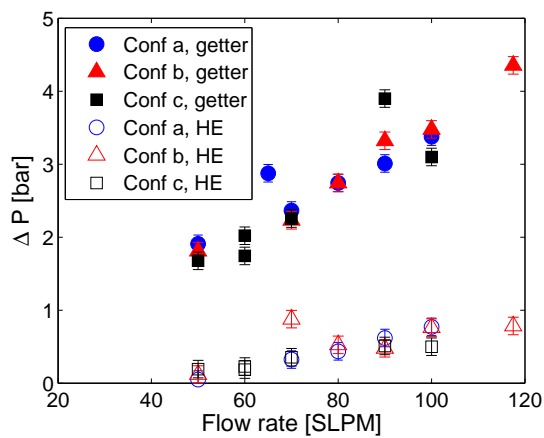}
\caption{Pressure drop as a function of flow rate, measured across the getter (filled markers) and the HE  - incoming gas pressure minus chamber pressure (unfilled markers), in three HE configurations a, b and c, as explained in the text.}
\label{fig:pressures}
\end{center}
\end{figure}

{\it Pressure drops and dynamical resistance}: We have measured the pressure drops across the getter and the HE, which proved to be the main sources of dynamical resistance to the flow. Figure \ref{fig:pressures} shows the pressure drops as a function of the flow rate, on the getter and on the HE. The results are shown for the three configurations of HEs. The pressure drop across the different components grows as a function of the flow rate. At a Xe vapor pressure of $ >2$ bar in the detector, a pressure of $ \sim 8 $ bar on the outlet of the pump is required for driving a constant flow of $ \sim 120 $ SLPM.

{\it Buffer volume}: At flow rates larger than $ 50 $~SLPM, the output pressure of the pump starts fluctuating with a high frequency (30 Hz, half the AC frequency supplied to the pump) and an amplitude that exceeds 1.5 bar at flow rates above 100 SLPM. To prevent damage to the system and provide more stability to the flow we  added a buffer volume of $ \sim 1 $~liter, close to the outlet of the pump. The displacement volume of the diaphragm pump is about 50~cm$^3$, so the buffer volume decreases the typical pressure change by a factor $\sim 20$. The buffer volume proved to be efficient in damping the pressure fluctuations, as expected.

{\it Increasing the maximum flow rate}: The factor limiting the flow rate in our configuration is the pressure drop across the MFC, since the flow  forced by the recirculation pump is most sensitive to the inlet pressure. This drop (10-15 psi, \cite{MFC_man}) is required for the flow measurement. The control solenoid valve itself has a lower resistance to the flow. In order to be able to reach high flow rates we increased the pressure in the chamber by setting a slightly higher temperature set point on the cold finger. The chamber pressure required for a flow of 120~SLPM with the two HEs setup (configuration c) was 2.56 bar absolute.
In future tests we plan to separate the flow metering and the controlled valve, and place the valve at the inlet of the pump and the measurement device at the outlet. This will allow a faster gas flow without a strong restriction at the outlet. However, as mentioned earlier, the pressure at the outlet of the pump will increase, requiring a different pump or parallel getters.

In future tests, a different type of pump will be tested \cite{Qdrive}, which employs an acoustic compression
technique, for recirculating the Xe gas. This technology  has a potentially lower leak rate and a lower $^{222}$Rn emanation rate. A prototype that will allow for tests with the system has been constructed and will be available soon.  

{\it Bypassing the getter}: Changing the recirculation flow rate changes the steady state pressure of the detector, due to the different dynamics and heat input. We found that at high flow rates, bypassing the getter increases the pressure significantly, where the increase in pressure is more pronounced at higher flow rates. For instance, at 70~SLPM the bypassed steady state pressure is $ \sim 2.7$ barg, while for the same conditions, only flowing through the getter, the steady state pressure is slightly below 1 barg. We believe that the reason for this behavior is that the gas flowing through the getter is cooled by a radiator on the way out of the getter, whereas bypassing the getter takes the gas from the outlet of the pump almost directly back into the HE. At high flow rates the temperature of the gas at the outlet of the pump is very high, up to a few hundred degrees C.

\subsection{Heat Exchange Efficiency}
The cooling power of the PTR is measured by keeping the detector under vacuum and letting the cold head reach the temperature set point, corresponding to the desired LXe operation temperature. The power supplied by the heaters around the cold finger compensates the cooling power, keeping the temperature constant. Allowing the system to reach a steady state by leaving it cooling for about 12~h we measure directly the current and voltage applied to the heaters, thus finding the power that exactly cancels the PTR cooling. This value, in general, depends on the set point temperature, and was measured to be 208 W at a temperature of 173~K on the cold finger. 

The efficiency $ \varepsilon $ of a HE is defined as the fraction of heat, required for vaporization and temperature change, that is kept outside the system. For the change from LXe at 2 bar absolute, on the phase change line (177.9~K), to room temperature GXe we  can write

\begin{equation}
\varepsilon(r)=1-\frac{P(0)-P(r)-R(r)}{10.74\times r},
\end{equation}
where $ r $ is the flow rate [SLPM], P(r) is the heater power [W] at a given circulation rate $ r $, and $ R $ is a term that accounts for extra heat that leaks into the system around the HE, which is at room temperature when there is no flow. When using super insulation $ R $ can usually be neglected, as is done here. The value 10.74 W/SLPM corresponds to an isobaric vaporization of LXe at 2 bar absolute followed by an increase in temperature from 177.9~K to 293.15~K, of which 8.88 W/SLPM is spent on the enthalpy change during evaporation \cite{Lemmon2011:ni}.

\begin{figure*}[ht]
\begin{minipage}[t]{0.5\linewidth}
\centering
\includegraphics[width=6cm]{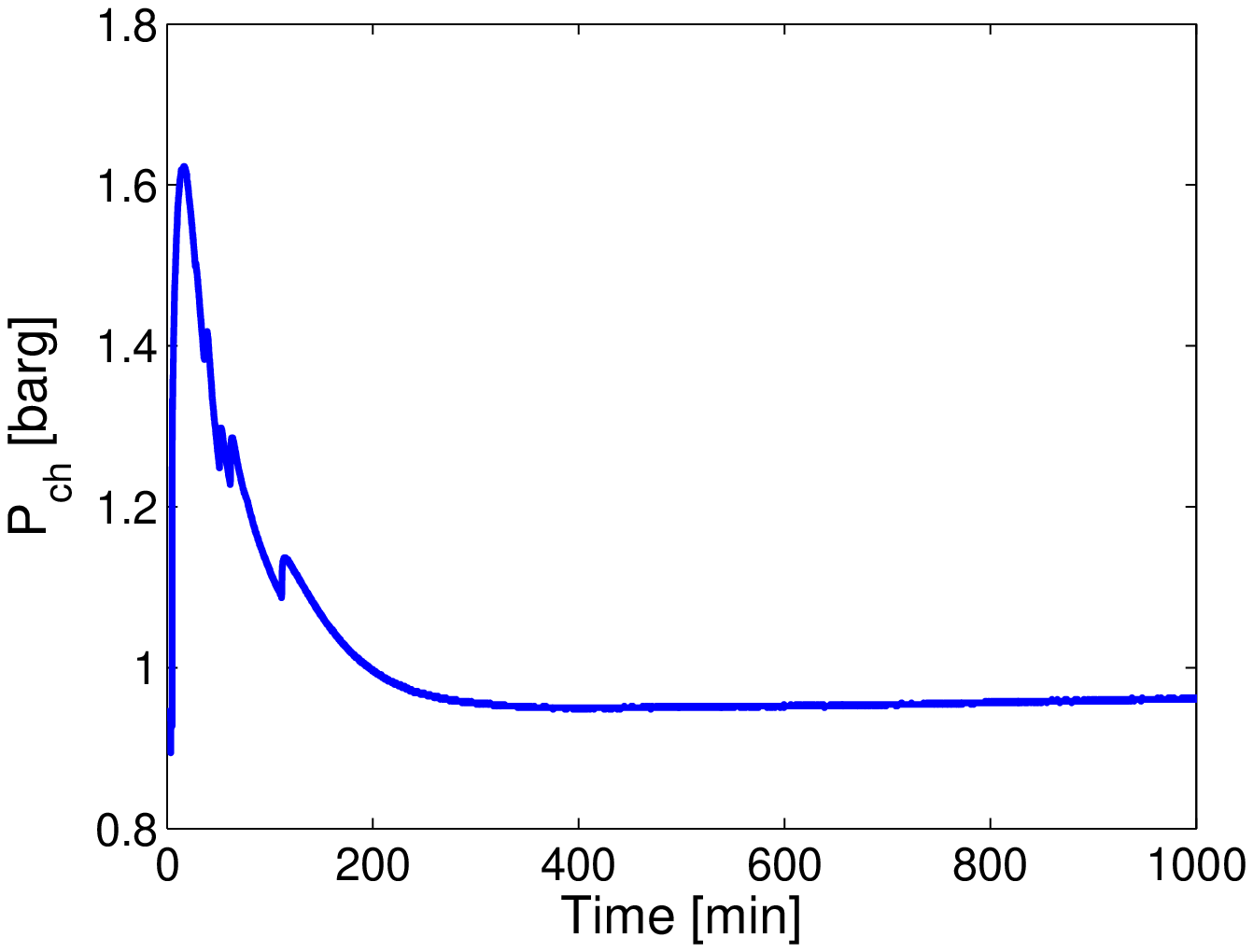}\caption{Pressure inside the detector as a function of time, following the initial filling and starting of circulation at a rate of 50~SLPM.}
\label{fig:p_stab_start}
\end{minipage}
\hspace{0.5cm}
\begin{minipage}[t]{0.5\linewidth}
\centering
\includegraphics[width=6cm]{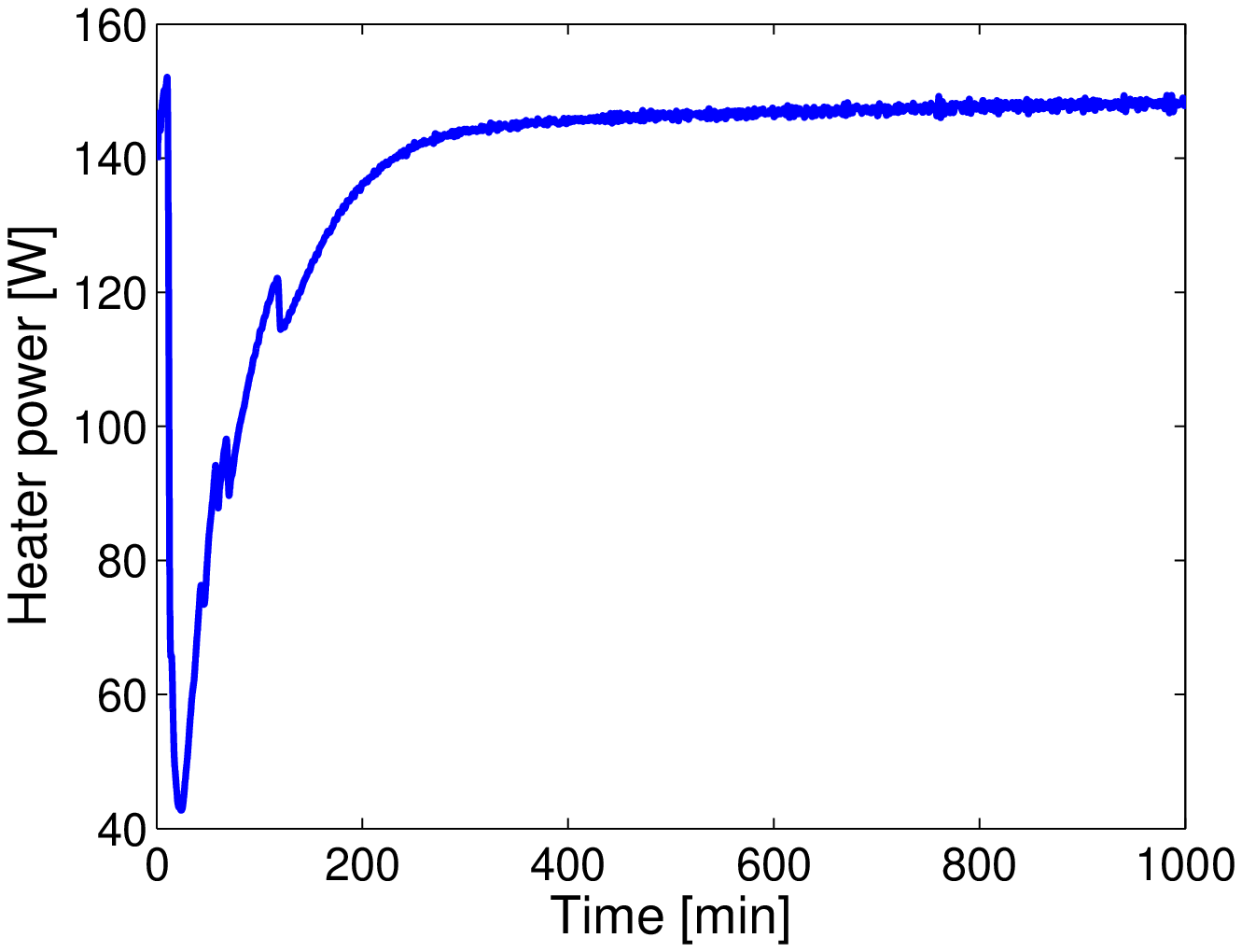}\caption{Heater power as a function of time, following the initial filling and starting of circulation at a rate of 50~SLPM.}
\end{minipage}
\end{figure*}

\begin{figure*}[ht]
\begin{minipage}[t]{0.5\linewidth}
\centering
\includegraphics[width=6cm]{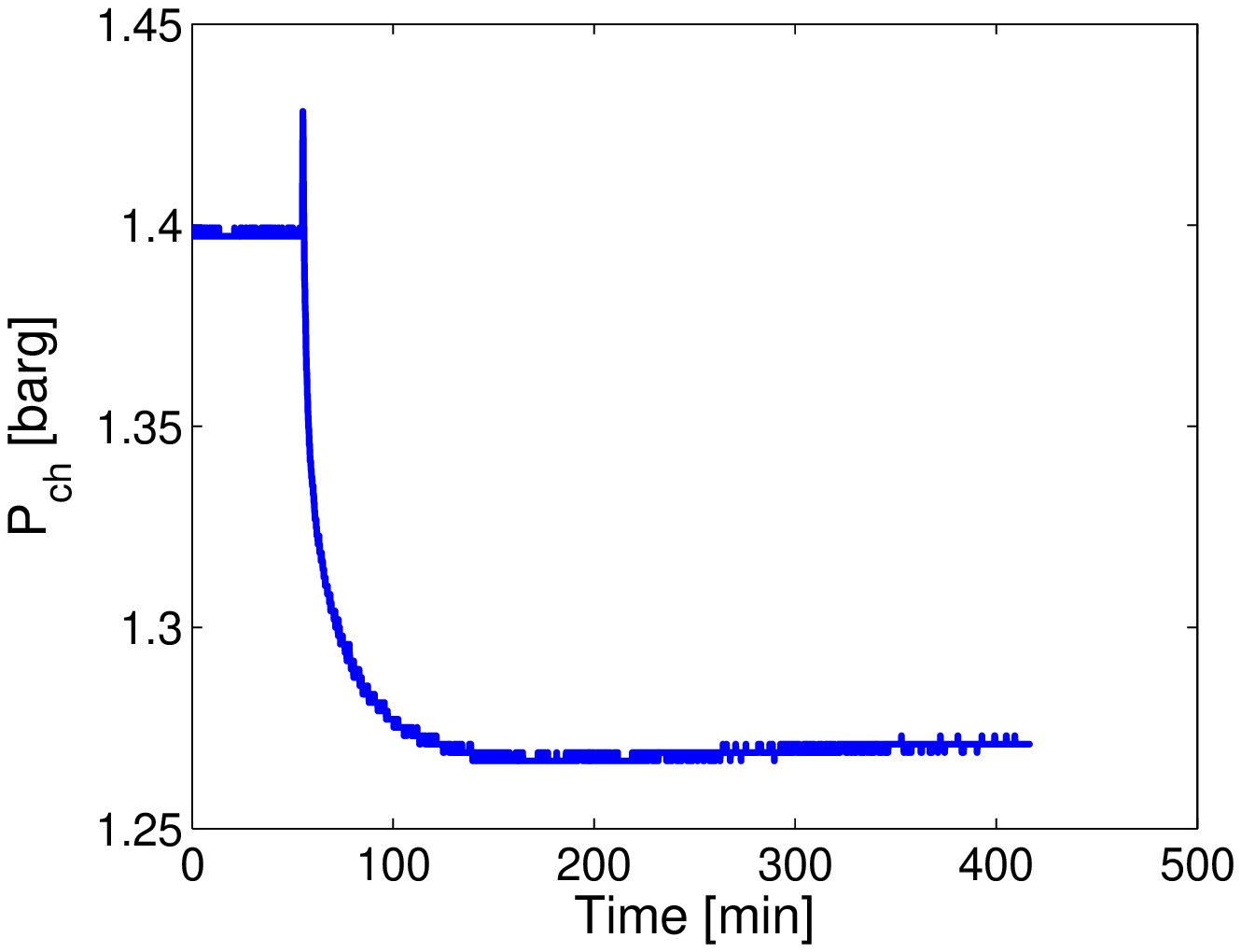}\caption{Pressure inside the detector as a function of time, following a change of circulation rate from 100~SLPM to 60~SLPM.}
\end{minipage}
\hspace{0.5cm}
\begin{minipage}[t]{0.5\linewidth}
\centering
\includegraphics[width=6cm]{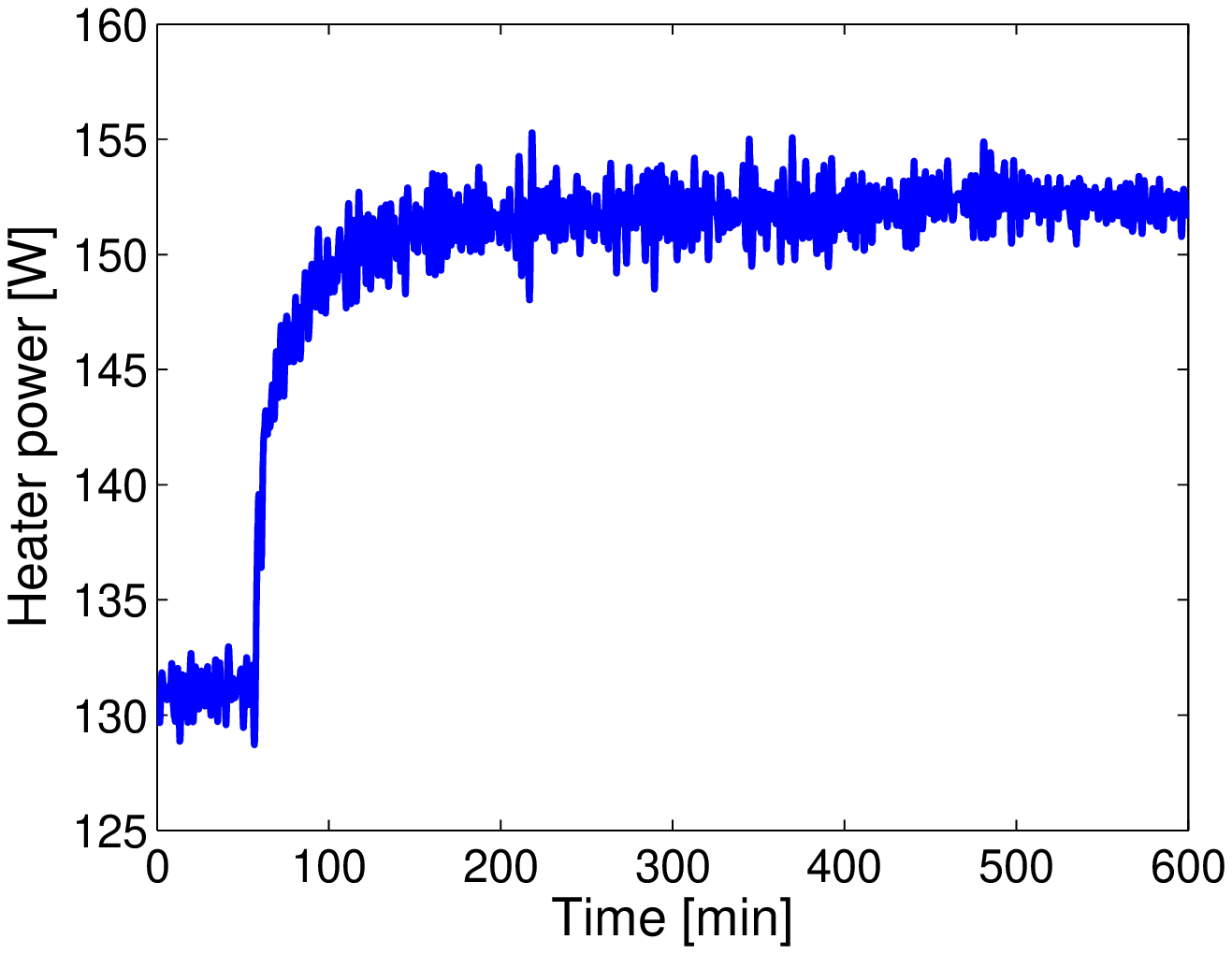}\caption{Heater power as a function of time, following a change of circulation rate from 100~SLPM to 60~SLPM.}
\label{fig:pow_stab_change}
\end{minipage}
\end{figure*}

Proper measurements of the heater power require that the system is sufficiently close to a thermal steady state. After the initial filling and start of circulation, the relaxation time is of the order of 12~h. A change of circulation speed also changes the thermal and pressure balance, and the relaxation time from one steady state to another is typically a few hours. Figures \ref{fig:p_stab_start} to \ref{fig:pow_stab_change} show the typical evolution of the detector pressure and heater power after the initial filling and change of circulation rate.

{\it Small HE}:  Initially, measurements with the small HE ($ \sim 0.5 $ l volume) were carried out. At flow rates up to ~20 SLPM the HE efficiency was measured to be in the range of 90\%-95\%. Since the system was designed for high flow rate, its sensitivity is limited when circulating at low flow rate, hence the large error. This result is compatible with that previously reported in \cite{Giboni11}. At higher flow rate, 40 SLPM, the efficiency drops to $  \sim$86\%.
The maximum rate attainable was 48 SLPM, at which liquid Xenon started spilling out of the HE, into the pipes of the gas system (observed as a violent pressure increase on the inlet of the pump).  At that point the heater power was still not zero, meaning there was enough cooling power to handle higher circulation speeds at that efficiency. When filling more LXe into the detector,  the maximum flow  was reduced to 35 SLPM. This points to the heat exchange role  played by the pipe carrying the liquid through the Xe gas phase, from the detector to the HE. A higher liquid level inside the chamber reduces the internal heat exchange, that takes place when the LXe in the tube taking the liquid exchanges heat with the gas phase in the  detector before reaching the HE. From these measurements we conclude that at moderate and high circulation speeds the HE is partially filled with liquid, which boils off absorbing the latent heat from the incoming gas. The liquid level inside the HE increases as a function of the circulation rate.

\begin{figure}[htb]
\begin{center}
\includegraphics[width=8cm]{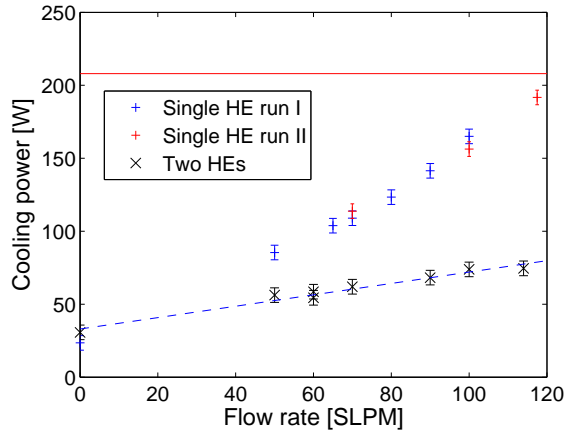}
\caption{The cooling power required for circulation at different flow rates and HE configurations. The red solid line at 208 W represents the maximum cooling power available, and the blue dashed line is a linear fit to the two HE configuration points, with a slope of 0.39 W/SLPM (corresponding to $>$96\% efficiency).}
\label{fig:power}
\end{center}
\end{figure}

{\it Large HE}: Following the above measurements, a larger HE was used, with 60 plates and volume of about 3.8~l. With this setup, circulation speeds of up to 120~SLPM were reached, with a margin of about 10 W cooling power remaining. The cooling power required as a function of circulation rate is shown in Figure \ref{fig:power}. The inferred heat exchange efficiency with this configuration is close to 90\% at flow rates up to 120 SLPM.

{\it  HEs in series}: Using the two HEs in series proved to be much more efficient than a single unit. The improvement  is significantly larger than that expected from the increase in volume of the combined units. Figure \ref{fig:power} also shows the required cooling power as a function of flow rate in this configuration, up to a rate of 114 SLPM. At that flow rate, the heaters still supply a power of about 130~W, which translates to an efficiency $ \varepsilon \approx 96\% $.

\subsection{About the physics of heat exchange}
The heat exchange involves two main processes: the phase transition and the gas temperature change.

{\it Phase transition:} Since about 80\% of the heat goes into latent heat of the gas-liquid phase transition, it is important to understand the way this heat is transferred between the ingoing and outgoing Xe. The key for efficient heat exchange is a temperature difference $ \Delta T_{ph} $ that drives the heat transfer between the two Xe flows. This  is the difference between the condensation temperature of the incoming warm Xe and the boiling temperature of the outgoing cold Xe,  $\Delta T_{ph}= T_{gl}(P_i)-T_{gl}(P_o)$, where $ T_{gl}(P) $ is the pressure dependent gas-liquid phase transition temperature. For dynamical gas flow reasons, and as shown in section \ref{subsec:gas_circ}, $ P_i>P_o $, and since $ T_{gl}(P) $ is an increasing function, $\Delta T_{ph}>0$ which leads to an effective heat transfer.

{\it Gas temperature change:} Once the LXe coming out of the detector is evaporated, it goes through the HE, thermally coupled to the incoming Xe gas through the heat conducting metal plates. The  difference between the initial gas temperatures is $\sim$120~K. Let us assume a very simplistic model for gas heat exchange: Two streams of Xe gas coupled through $ n $ plates, each of width $ W $ and length $L$, in the $ x $ direction from $ x=0 $, so that $ T_1(0)=T_2(L)+\Delta T=T_h $. We designate the plates' thickness as $ d $ and their heat conductivity as $ \kappa $. The density of Xe gas and its heat capacity are $ \rho $ and $ C_v $, respectively. Neglecting the heat capacity of the plates and any temperature gradient in the gas perpendicular to the plates we can write for a gas flow rate of $ r $ (mass per unit time):
\begin{equation}\label{eq:HE1}
T_1'=\frac{dT_1}{dx}=-\alpha (T_1(x)-T_2(x))
\end{equation}
\begin{equation}\label{eq:HE2}
T_2'=\frac{dT_2}{dx}=-\alpha (T_1(x)-T_2(x))
\end{equation}
where
\begin{equation}
\alpha \propto \frac{\kappa \rho nW}{rC_v d}.\label{eq:alpha}
\end{equation}
Equation \ref{eq:alpha} is an equality under the assumption of perfect heat transfer between the gas and the plates, as well as perfect heat transfer inside the gas perpendicular to the flow direction. The proportionality constant turns out to be much smaller than 1, mostly due to inefficiencies in transferring the heat to the bulk of the gas. Equations \ref{eq:HE1} and \ref{eq:HE2} lead to an efficiency (for the gas heat exchange only) of
\begin{equation}
\varepsilon_{gas}=\frac{\alpha L}{1+\alpha L}.
\end{equation}
If $h$, the space between plates, is constant, then the figure of merit for heat exchange efficiency is the volume of the HE (equals $nhWL$) divided by the flow rate $r$.

It appears that in the case of liquid-gas heat exchange the description of the gas heat exchange is more complicated. The striking improvement of efficiency when adding in series the smaller HE is not expected by its volume, which is $ \sim 14\% $ of the larger HE's volume. The difference lies, in our opinion, in the temperature of the HE itself. When circulating at high rates, part of the bottom HE is filled with LXe, as established by the tests with the small HE. This amount of liquid increases as the rate increases, and the liquid cools down the metal plates above the liquid level, thus preventing efficient heat exchange of the Xe gas on top of the LXe. The use of a second HE, thermally decoupled from the bottom one, allows the gas to exchange heat efficiently thus decreasing the total energy loss by a factor of almost 3.

\section{Conclusions}
We have carried out measurements to study and demonstrate the ability to flow Xe gas at high rates, above 100~SLPM, for applications in future detectors such as XENON1T.
We have found that a high flow rate of Xe requires relatively high absolute pressures and pressure gradients that can reach 8 bar at a flow above 100~SLPM. With the use of a high capacity pump and $1/2''$ tubing for the gas system, we have shown that the getter is the main restriction on the gas flow. We observed that a buffer volume on the outlet of the circulation pump is necessary to avoid large pressure fluctuations. We also note that a bypass of the getter increases significantly the pressure in the LXe detector, probably due to the high temperature of the Xe gas after being compressed by the circulation pump. 
We have shown that the high flow rate  with heat exchange requires that there be LXe inside the HE itself. This amount is non negligible (estimated to be $>1$~kg at 45~SLPM in our system), and influences the efficiency of the heat exchange. A single parallel plate HE, even as large as 4 l, is still limited to $\sim90\%$ efficiency. We find that using two HEs in series increases the efficiency to $\sim96\%$, due to the decoupling of the top HE from the LXe that exists inside the bottom one. This efficiency is consistent with that projected from tests carried out at  much lower circulation rates and previously reported by our group. 

Acknowledgements: 
This work was supported by the NSF with an award to the Columbia Astrophysics Laboratory (PHY-1047794) for the project entitled R\&D of LXe Detector technology for dark matter experiments. We also acknowledge support from the Weizmann Institute of Science with a Research Fellowship to R.~Budnik  (WIS CU10-1945).

\end{document}